\documentclass[a4paper,11pt]{article}
\usepackage{pos}

\usepackage{braket}
\usepackage{bm,bbm}

\newcommand\beq{ \begin{eqnarray} }
\newcommand\eeq{ \end{eqnarray} }

\title{New computational methods in lattice gauge theory
- quantum computation and tensor networks -}
\ShortTitle{New computational methods in lattice gauge theory}

\author*[a,b]{Etsuko Itou}

\affiliation[a]{Yukawa Institute for Theoretical Physics, Kyoto University\\
Kitashirakawa Oiwakecho, Sakyo-ku, Kyoto 606-8502 Japan}

\affiliation[b]{RIKEN Center for Interdisciplinary Theoretical and Mathematical Sciences (iTHEMS), RIKEN\\
2-1 Hirosawa, Wako, Saitama 351-0198 Japan}

\emailAdd{itou@yukawa.kyoto-u.ac.jp}

\abstract{Lattice QCD calculations have been conducted using large-scale classical computers based on the Lagrangian formalism of field theory for the past 40 years. On the other hand, the advent of quantum computers has brought increasing attention to first-principles computational methods based on the Hamiltonian formalism compatible with these machines. In this talk, we discuss recent results on how to calculate hadronic properties using the Hamiltonian formalism. }

\FullConference{The 21st International Conference on Hadron Spectroscopy and Structure (HADRON2025)\\
 27 - 31 March, 2025\\
Osaka University, Japan\\}

\begin{document}
\maketitle

\section{Introduction}
In recent years, new ab initio calculation methods of gauge theory have attracted attention to formulations that write down the theory in a Hamiltonian formalism, especially using spin Hamiltonians to fit quantum computation and tensor networks.
J.~B.~Kogut and L.~Susskind proposed the Hamiltonian formulation of lattice gauge theory in 1975~\cite{Kogut:1974ag}. It was the year after Wilson's 1974 paper on lattice gauge theory~\cite{Wilson:1974sk}, but so far, the numerical calculation has been done by the Euclidean path integral method based on the Lagrangian formalism as Lattice Monte Carlo QCD.
Ab initio calculation by the operator formalism based on the Hamiltonian has not made much progress for the past 40 years.

On the other hand, the conventional Lattice Monte Carlo method has been suffering from serious problems, namely the sign problem, for many years. Typically, it arises in three situations related to the physics of QCD; the real-time evolution, the finite-density QCD, and QCD (or Yang-Mills theory) with the topological $\theta$-term.
The sign problem arises from the use of an importance-sampling method that uses the Euclidean action as Boltzmann weights in the Monte Carlo method to do the Euclidean path integral efficiently, so this would not arise if a different algorithm were used.
In particular, the Hamiltonian formulation has the advantage that it can naturally formulate real-time evolution.

The most significant difficulty in calculating the Hamiltonian formulation of gauge theories is that the Hilbert space of the gauge field, which is a boson, is infinite-dimensional. Forcing a truncation of this space explicitly breaks gauge symmetry. Currently, various proposals have been made on how to avoid this problem.
On the other hand, it is known that under certain special conditions in low dimensions, the gauge degrees of freedom can be written entirely in terms of fermion degrees of freedom. A notable example is 1+1-dimensional quantum electrodynamics (QED).
This theory is also known as the Schwinger model. The Schwinger model has properties similar to those of QCD, such as confinement (due to the Coulomb force) and the (discrete) chiral symmetry breaking.
Regarding the elimination of gauge degrees of freedom, this is related to the fact that the 1+1-dimensional gauge field inherently lacks transverse components. Technically, by fixing a gauge, solving Gauss's law, and imposing open boundary conditions, the degrees of freedom of all gauge fields can be expressed in terms of fermion degrees of freedom.

Now, one interesting point about first-principles calculations using the Hamilton formalism is that they allow physical quantities to be investigated in different ways from the conventional ones.
For example, when calculating the $q$-$\bar{q}$ potential, conventional Lattice QCD calculates the potential $V(r)$ by calculating the Wilson loop and reads the exponential part of the $\langle W( r\times T) \rangle \sim e^{-TV(r)}$ in the large $T$ limit. Reading the exponential part tends to result in noisy data, and techniques such as smearing are necessary to obtain clear signals.
On the other hand, since the Hamiltonian formalism allows the energy of the system to be calculated directly, it is possible to calculate the potential directly by preparing a Hamiltonian that represents a situation where $q$ and $\bar{q}$ with distance $r$ are placed as probes. Such a calculation has been done using the quantum algorithm by introducing a site-dependent topological $\theta$-term for the 1-flavor Schwinger model (see e.g., Refs.~\cite{Honda:2021aum, Honda:2021ovk}).

In this manuscript, we focus on how the mass spectrum of composite particles, i.e., hadrons, can be calculated using the Hamilton formalism. We demonstrate a numerical simulation based on the tensor network method for the 2-flavor Schwinger model.
The main results are originally obtained in Refs.~\cite{Itou:2023img, Itou:2024psm}

\section{2-flavor Schwinger model}
Let us consider the 2-flavor Schwinger model, which is a $(1+1)$-dimensional quantum
electrodynamics with two species of Dirac fermion.
The Lagrangian density with the Minkowski metric
$\eta_{\mu\nu}=\mathrm{diag}(1,-1)$ is given by 
\begin{equation}
\mathcal{L}=-\frac{1}{4g^2}F_{\mu\nu}F^{\mu\nu}+\frac{\theta}{4\pi}\epsilon_{\mu\nu}F^{\mu\nu}+\sum_{f=1}^{N_{f}}\left[i\bar{\psi}_{f}\gamma^{\mu}\left(\partial_{\mu}+i A_{\mu}\right)\psi_{f}-m\bar{\psi}_{f}\psi_{f}\right],
\label{eq:Lagrangian}
\end{equation}
where $F_{\mu\nu}=\partial_{\mu}A_{\nu}-\partial_{\nu}A_{\mu}$ is
the field strength, $g$ is the gauge coupling, and $\theta$ is the vacuum angle describing the background electric flux. 
The index $f$ labels the flavor.
We set that the two fermions have a degenerate mass $m$.
Note that, here we add the $\theta$-term into the Lagrangian. Thus, the corresponding Lattice Monte Carlo calculation for this model suffers from the sign problem.

Here, we give the lattice Hamiltonian formalism for numerical computation. 
We consider the 2-flavor Schwinger model on an open interval.
In this case, the Gauss' law condition can be solved explicitly, 
so that the Hamiltonian is described only by fermions after the gauge fixing.
We adopt the staggered fermion~\cite{Kogut:1974ag, Susskind:1976jm} for the lattice discretization and apply the Jordan-Wigner transformation to obtain the spin Hamiltonian with a finite-dimensional Hilbert space,
\begin{align}
    H= & \frac{g^2a}{8}\sum_{n=0}^{N-2}
    \left[\sum_{f=1}^{N_{f}}\sum_{k=0}^{n}\sigma_{f,k}^{z}
    +N_{f}\frac{(-1)^{n}+1}{2}+\frac{\theta}{\pi}\right]^2 \nonumber \\
     & -\frac{i}{2a}\sum_{n=0}^{N-2}\left(
    \sigma_{1,n}^{+}\sigma_{2,n}^{z}\sigma_{1,n+1}^{-}
    +\sigma_{2,n}^{+}\sigma_{1,n+1}^{z}\sigma_{2,n+1}^{-}
    -\mathrm{h.c.}\right)
    +\frac{m_{\mathrm{lat}}}{2}\sum_{f=1}^{N_f}\sum_{n=0}^{N-1}(-1)^{n}\sigma_{f,n}^z.
    \label{eq_H_spin}
\end{align}
Here, the lattice fermion mass $m_{\mathrm{lat}}$ and the mass $m$ of the continuum theory are related as $m_{\mathrm{lat}}:=m-\frac{N_{f}g^{2}a}{8}$~\cite{Dempsey:2022nys}.

 According to semiclassical analyses using the bosonization technique by R.~F.~Dashen~\cite{Dashen:1975hd} and S.~R.~Coleman~\cite{Coleman:1976uz}, the low-lying states are expressed by the following composite states, namely ``hadrons" or ``mesons", at $\theta=0$. 
\begin{alignat}{3}
    &\rm{pion: }~~ &&\pi_a=-i \bar{\psi}\gamma^5 \tau_a \psi \quad  && (J^{PG}=1^{-+}),
    \label{eq:pi_meson} 
    \\
    &\rm{sigma~meson: }~~&& \sigma=\bar{\psi}\psi && (J^{PG}=0^{++}),
    \label{eq:sigma_meson}
    \\
    & \rm{eta~meson: }~~&& \eta=-i \bar{\psi}\gamma^5\psi && (J^{PG}=0^{--}). 
    \label{eq:eta_meson}
\end{alignat}
Here, $J^{PG}$ shows their isospin, parity, and $G$-parity quantum numbers.
At $\theta \ne 0$, there is a mixing between the scalar and pseudo-scalar operators; therefore, the mass eigenstate of each composite particle becomes
\begin{equation}
    \pi_{a} = -i\bar{\psi}\, e^{i\frac{\theta}{2}\gamma^5} \gamma^{5} \tau_{a} \psi,
    \qquad
    \sigma = \bar{\psi}\, e^{i\left(\frac{\theta}{2}+\omega(\theta)\right)\gamma^5} \psi,
    \qquad
    \eta = -i\bar{\psi}\, e^{i\left(\frac{\theta}{2}+\omega(\theta)\right)\gamma^5} \gamma^{5} \psi.
    \label{eq_axial_rot}
\end{equation}
Here, the extra rotation $\omega(\theta)$ comes from the effect of the $\sigma-\eta$ mixing.
Since the exact mixing angle is not known, it must be determined numerically.

\section{Calculation methods for "hadron" spectra in Hamiltonian formula }
In our work~\cite{Itou:2023img}, we proposed three calculation methods for the mass spectra of composite states:
\begin{enumerate}
    \item Correlation-function scheme

It is essentially the same as the conventional one (point-point correlation function): We measure the spatial correlation function of the composite operators, and from its long-distance behavior, we extract the effective mass of the low-lying state. 
 In Ref.~\cite{Itou:2023img}, we found that there is a power correction in $1+1$ d system, and the spatial connected two-point function is well-fitted by the Yukawa-type functional shape,
\beq
\langle \mathcal{O}(x) \mathcal{O}(0) \rangle_{\rm conn.} \sim \frac{e^{-Mx}}{x^{\alpha}},
\eeq
where $M$ and $\alpha$ are fitting parameters.
Suppose that the Lorentz symmetry is restored in the continuum and thermodynamic limit. The exponent $M$ goes to an effective mass of the low-lying state by taking these double limits.

    \item One-point-function scheme 

The measurement observable in this scheme is the one-point function of the bulk composite operator.
In our simulation setup, we impose the open boundary condition, and then there is an edge mode at the boundaries. It gives some boundary state $| \rm{Bdry} \rangle$, and we can identify it as a wall source for the bulk operator.
The asymptotic behavior of the one-point function for the gapped system is
\beq
\langle \mathcal{O}(x) \rangle \sim e^{-Mx},
\eeq
where $M$ corresponds to the mass of the lightest state meson with the quantum number of the composite operator $\mathcal{O}$.

In this scheme, boundary conditions must be chosen appropriately to couple with states that have the desired quantum numbers. 
For instance, to access the mass of the pion, which corresponds to an isospin triplet state, at $\theta = 0$, we consider the theory at $\theta = 2\pi$, where the edge mode of the Symmetry-Protected Topological (SPT) phase can serve as a source for the triplet mesons~\cite{Itou:2023img}.

    \item Dispersion-relation scheme
    
In the Hamiltonian formula, we can measure the energy $E=\langle H \rangle$ and the momentum square $\langle \hat{K}^2 \rangle$ of the eigenstates of the Hamiltonian~\cite{Wall_2012, Banuls:2013jaa}. Furthermore, we can generate an $\ell$-th excited state as the lowest energy state for the modified Hamiltonian;
\begin{equation}
    H_{\ell} = H + W\sum_{\ell^{\prime}=0}^{\ell-1}\ket{\Psi_{\ell^{\prime}}}\bra{\Psi_{\ell^{\prime}}},
    \label{eq_H_ell}
\end{equation}
where we add a penalty term with a weight $W>0$ into the Hamiltonian Eq.~\eqref{eq_H_spin}. The penalty term gives an orthogonal condition with all lower states, from the ground state to the ($\ell-1$)-th excited state.
Identifying the species of the meson by measurement of the quantum numbers $J^{PG}$, the mass can be obtained using the dispersion relation, $E=\sqrt{M^2 + K^2}$~\footnote{See Ref.~\cite{Itou:2023img} for the explicit form of these operators written in the spin variables.}.
In our work, we impose the open boundary condition, so that the translational invariance is broken and the momentum operator does not commute with the Hamiltonian. Here, we evaluate the momentum square of the $\ell$-th excited state by subtracting the ground state contribution; $\Delta K^2_{\ell} \coloneq \langle \Psi_\ell | K^2 | \Psi_\ell \rangle - \langle \Psi_0 | K^2 |\Psi_0 \rangle $.

\end{enumerate}

In this manuscript, we mainly summarize the calculation results for the dispersion-relation scheme for $\theta=0$ and $\theta \ne 0$.

\section{Numerical results}
Our numerical calculation for this system written by the spin-Hamiltonian has been done using the density matrix renormalization group (DMRG) method~\cite{White:1992zz, White:1993zza} and the C++ library of ITensor~\cite{itensor}.
The DMRG is a variational algorithm based on the matrix product state (MPS). Thus, the wave function for the Hamiltonian is expressed by MPS.
In each step of the algorithm,
the matrices are updated to decrease the energy $E = \langle \Psi |H |\Psi \rangle$ as a cost function. 
In addition, we perform the low-rank approximation and thus the smaller singular values are
discarded, which amounts to an error $\Delta$. We determine the bound dimension by setting
the maximal bond dimension and also the cutoff parameter $\varepsilon$ on the error so that $\Delta \leq \varepsilon$.
Smaller $\varepsilon$ gives a better approximation, but also requires a larger bond dimension and increases the computational costs. 

In the simulations for the dispersion-relation scheme, the lattice size $N=100$, the lattice spacing $a=0.20$ and $\varepsilon=10^{-10}$, and the number of sweeps for the DMRG process $N_{\rm sweep} = 50, 80$.
For the other schemes, we utilize other values of $N,a, \varepsilon,$ and $N_{\rm sweep}$, but as we will show, the results are consistent with each other.
We conclude that using these setups, the discretization errors and finite-volume effects of our simulations are under control.

\subsection{$\theta =0$}
\begin{figure}[h]
    \centering
    \includegraphics[scale=0.4]{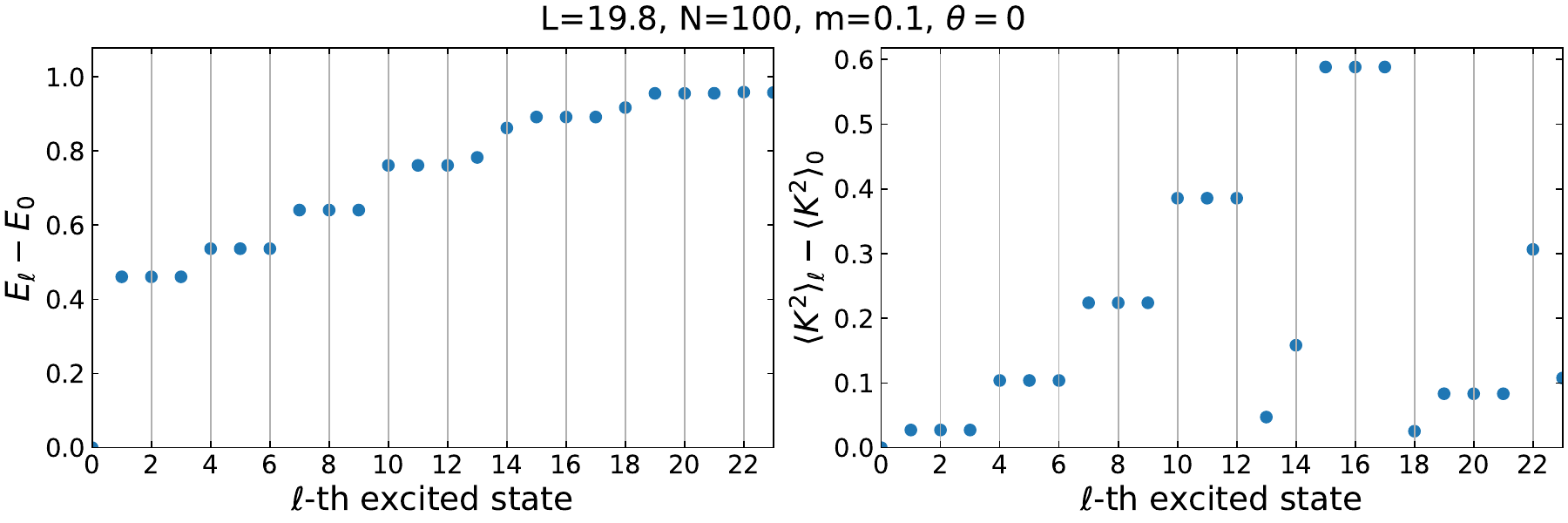}
    \caption{ The energy gap (left) and momentum square (right) of each $\ell$-th excited state.
    }\label{fig:E_and_K2_theta0}
\end{figure}

At $\theta =0$, we generated the MPS up to the 23rd excited state.
The energy gap $\Delta E = E_\ell - E_0$ and momentum square $\Delta K^2_\ell$ of each $\ell$-th excited state are shown in the left and right panels of Fig.~\ref{fig:E_and_K2_theta0}, respectively.
In this plot, we can see a clear structure of a triple degeneracy.

\begin{figure}[h]
    \centering
    \includegraphics[scale=0.14]{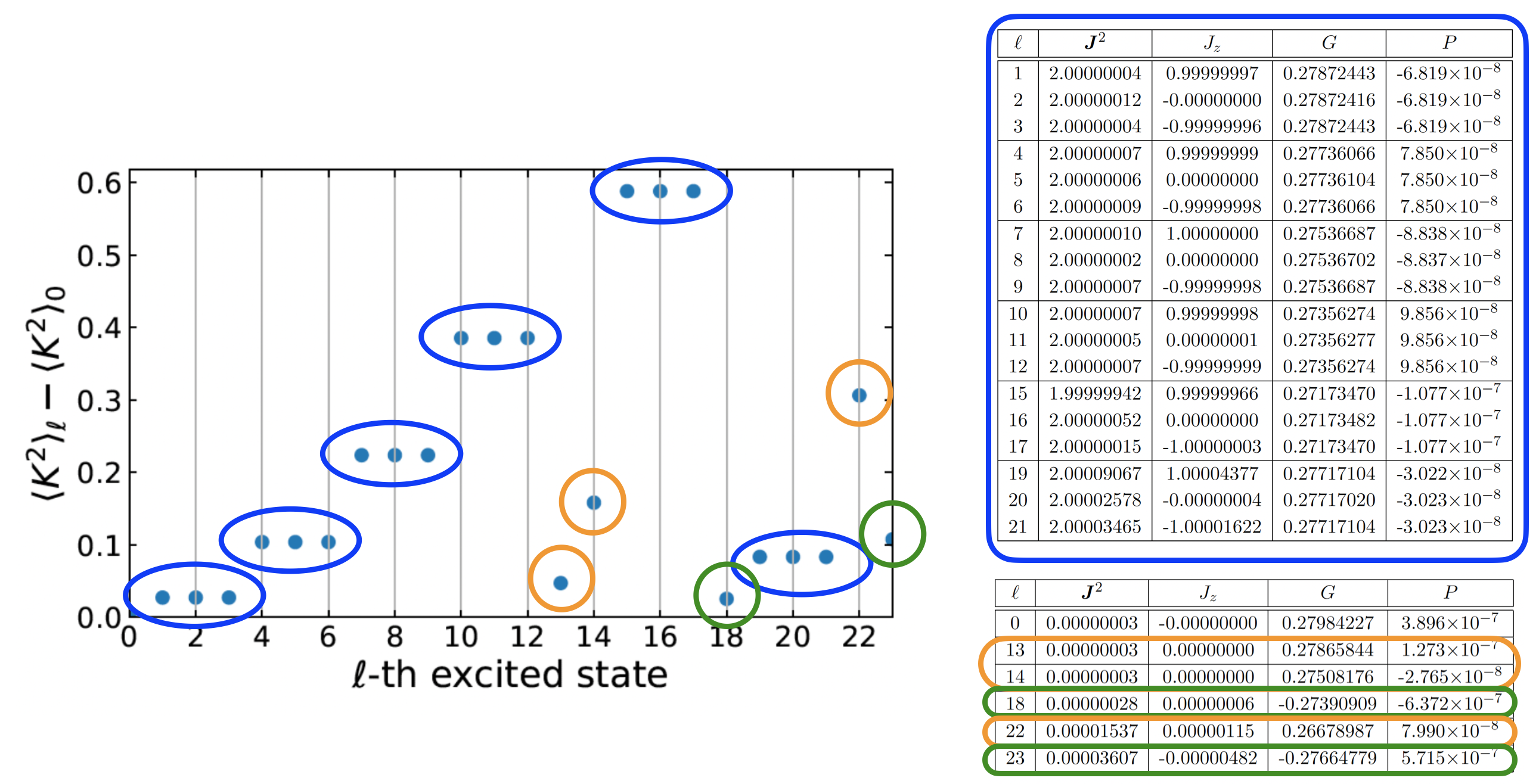}
    \caption{ Summary of the measurements of quantum numbers;  $\bm{J}^{2}, J_z, P,$ and $G$.
It shows that the states in the blue, orange, and green circles in both panels correspond to the states with $J^{PG}=1^{-+}$, $J^{PG}=0^{++}$, and $J^{PG}=0^{--}$, respectively.
    }\label{fig:quantum-number-theta0}
\end{figure}

To identify the states, we measure the expectation values of the isospin operators $\bm{J}^{2}, J_z$, the parity $P$, and the G-parity $G=Ce^{i\pi J_y}$ with the charge conjugation operator $C$.
The results are summarized in the right tables in Fig.~\ref{fig:quantum-number-theta0}.
Note that the even-odd determination of parity is based on the assumption that the lowest momentum state with the exact quantum numbers of the isospin and the G-parity is a zero mode for each meson.

\begin{figure}[h]
    \centering
    \includegraphics[scale=0.12]{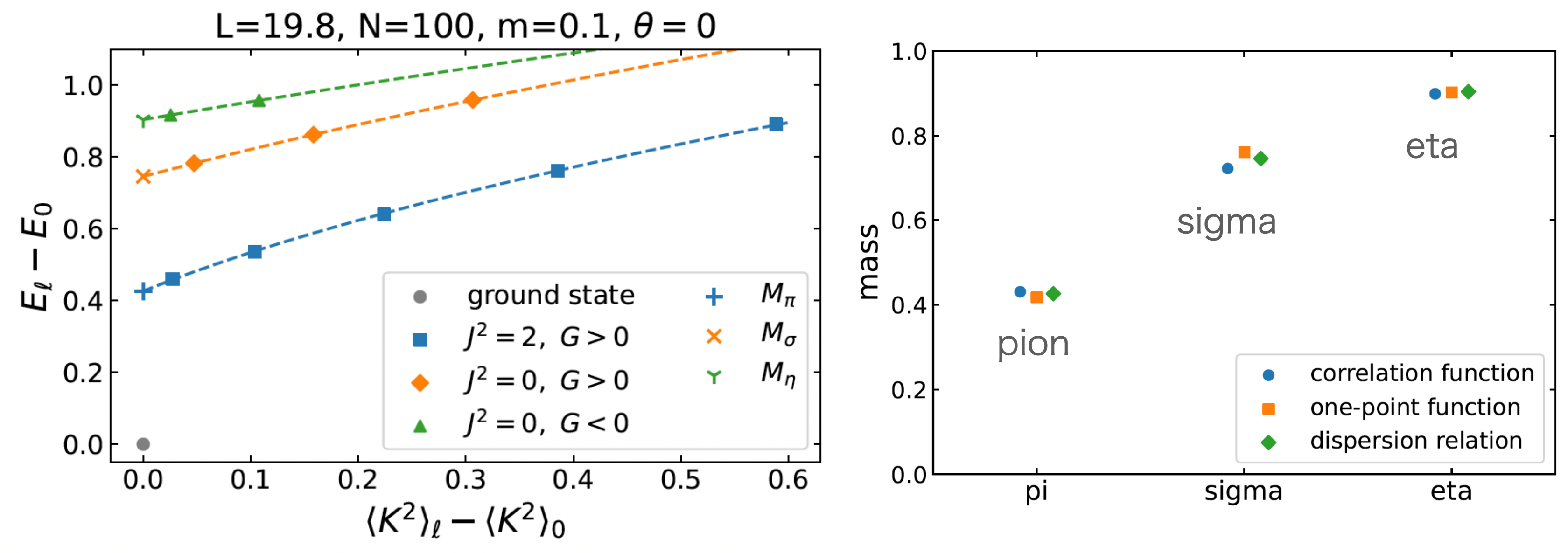}
    \caption{ 
    (Left) Fit results for the data of $\Delta E$ against $\Delta K^2_\ell$ for each meson state using the dispersion-relation ansatz (Right) Main results of Ref.~\cite{Itou:2023img}. Here we show the results obtained by three calculation methods for pion, sigma, and eta mesons.
    }\label{fig:result-theta0}
\end{figure}

After identifying the quantum numbers, we plot the energy gap against the momentum square and perform the fit of the data for each meson using the dispersion relation, $\Delta E = \sqrt{M^2 + b^2 \Delta K^2}$, as shown in the left panel of Fig.~\ref{fig:result-theta0}. Here, $M$ and $b$ are the fitting parameters. $b$ is introduced to improve the fit quality, and the fit results are almost consistent with $1$.

The fit result of $M$ can be regarded as the mass of each meson and is summarized in the right panel of Fig.~\ref{fig:result-theta0}.
This plot also shows the consistency of the three calculation methods.
The results obtained are consistent with the analytical predictions; the order of mass spectra ($m_{\pi}< m_{\sigma} < m_{\eta}$), the ratio of the pion and sigma ($m_{\sigma}/m_{\pi} = \sqrt{3}$), and the value of eta meson mass ($m_{\eta} \sim \sqrt{2/\pi}  g \sim 0.8 $).

\subsection{$\theta \ne 0$}
Now, let us move to the $\theta \ne 0$ regime, where there is the sign problem in the conventional Monte Carlo method.
Theoretically, it is known that some subtleties arise:
The operator mixing between the scalar and pseudo-scalar operators occurs. Thus, if we consider the correlation-function scheme and one-point function scheme, then we have to first determine the mixing angle and rotate the operator basis to the mass eigenstate. Furthermore, it is expected that the massive two-flavor Schwinger model at $\theta=\pi$ becomes almost a conformal field theory, namely SU$(2)$ level-1 Wess-Zumino-Witten model~\footnote{Exactly speaking, there is a tiny mass gap, $M_\pi \sim e^{-\# g^2/m^2}$~\cite{Dempsey:2023gib}.}.
Therefore, the functional form of the two-point or one-point function would be changed, so some modification is needed to confirm the conformality or the emergence of massless particles.

\begin{figure}[h]
    \centering
    \includegraphics[scale=0.12]{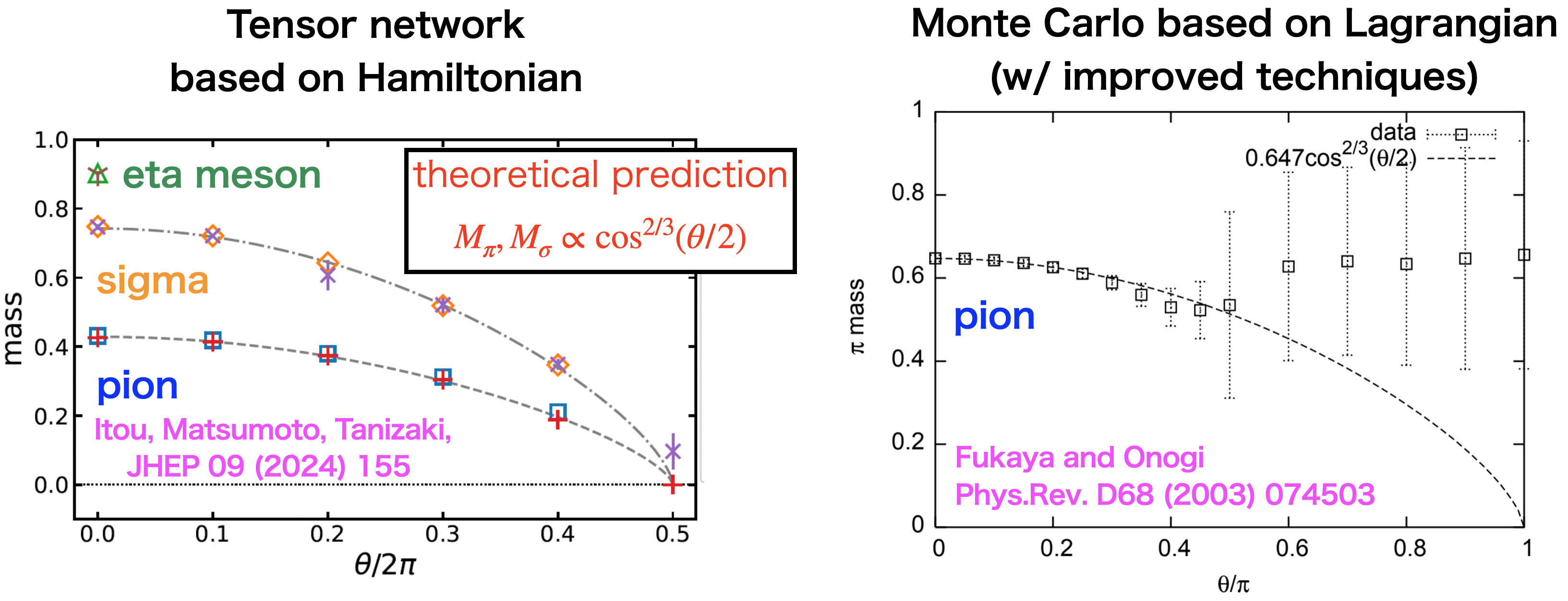}
    \caption{ 
    The $\theta$-dependence of mass spectra. (Left) Our results based on the Hamiltonian formula. The gray dashed and dotted curves denote $M_{\pi}(0) | \cos(\theta/2) |^{2/3}$ and $\sqrt{3} M_{\pi}(0) | \cos(\theta/2) |^{2/3}$, respectively.   (Right) Monte Carlo results based on the Lagrangian formula, which is originally given in Ref.~\cite{Fukaya:2003ph}. 
    }\label{fig:comp-result}
\end{figure}

On the other hand, the dispersion-relation scheme can be applied to the $\theta \ne 0 $ regime without modification.
\begin{figure}[h]
    \centering
    \includegraphics[scale=0.35]{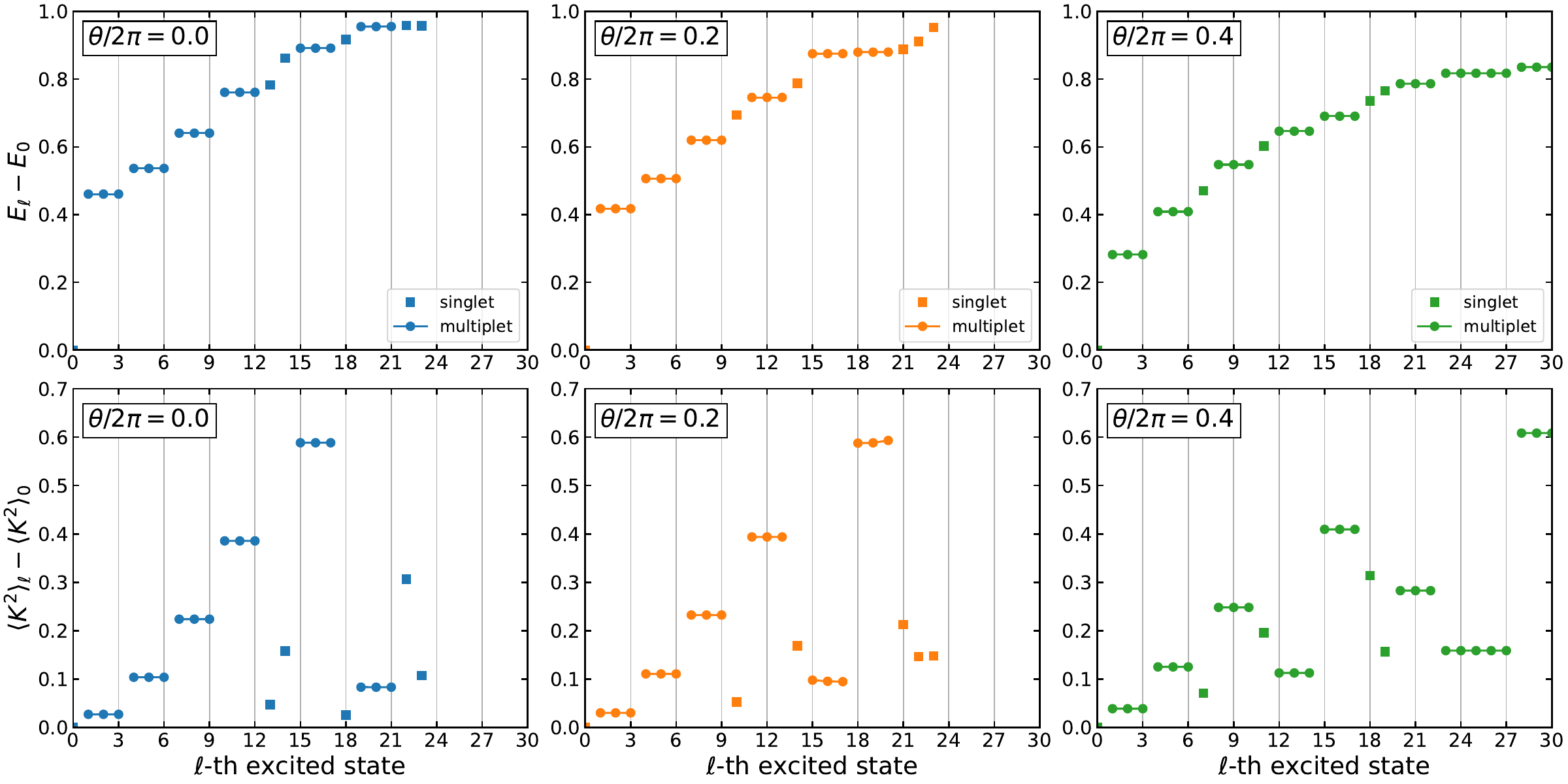}
    \caption{ 
   The energy gap (top panels) and momentum square (bottom panels) of each $\ell$-th excited state for $\theta/2\pi = 0$ (left), $0.2$ (middle), and $0.4$ (right), respectively.
    }\label{fig:E_and_K2_nonzero_theta}
\end{figure}
Figure~\ref{fig:E_and_K2_nonzero_theta} shows the energy gap (top panels) and momentum square (bottom panels) for $\theta/2\pi = 0$ (left), $0.2$ (middle), and $0.4$ (right), respectively.
Again, we can see several triplet structures in these plots.
\begin{figure}[h]
    \centering
    \includegraphics[scale=0.35]{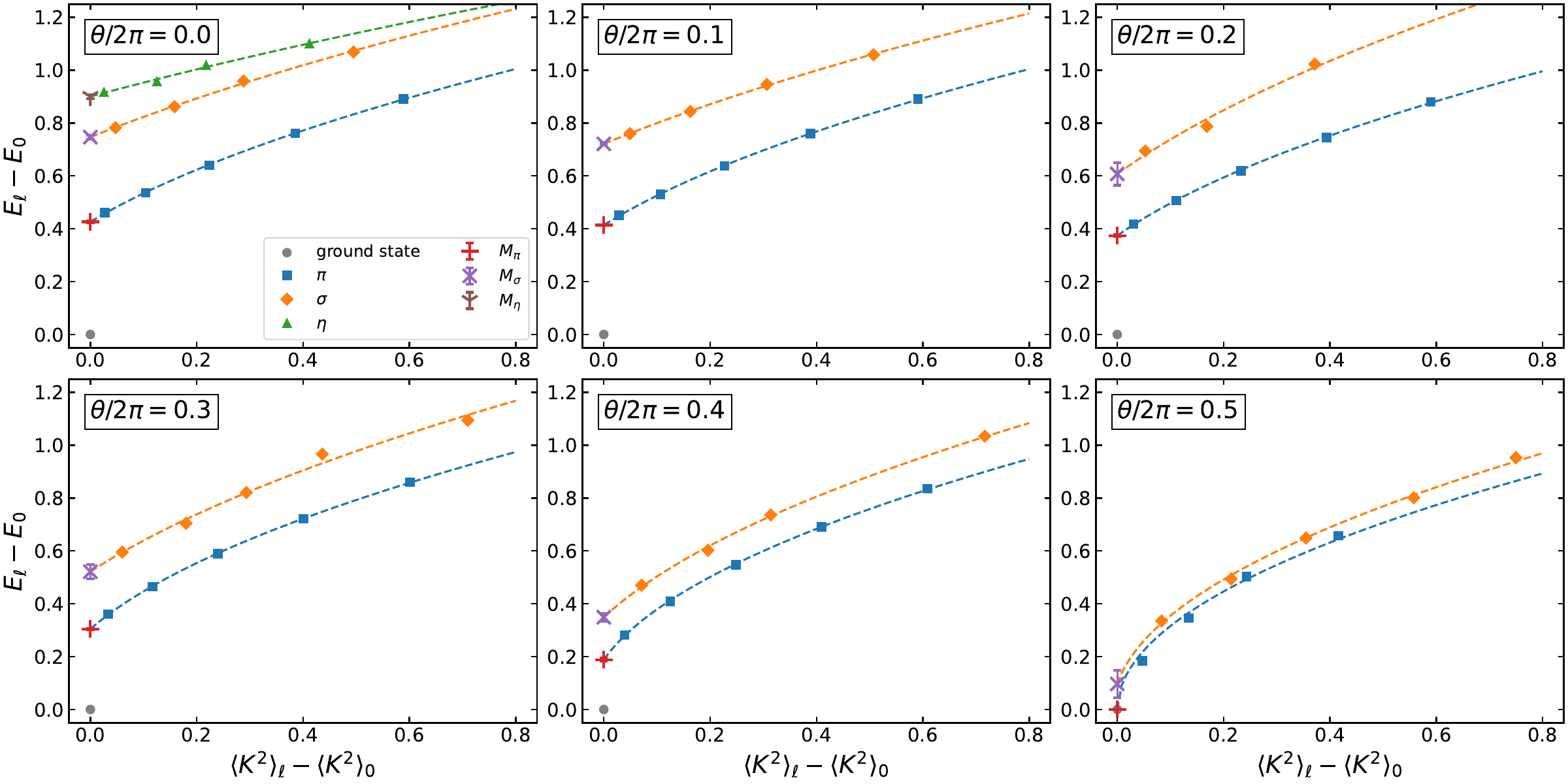}
    \caption{ 
   Fit results for the data of $\Delta E$ against $\Delta K^2_\ell$ for each meson state using the dispersion-relation ansatz for several values of $\theta$.
    }\label{fig:results-nonzero-theta}
\end{figure}
After identifying the quantum numbers, we plot the energy gap against the momentum square as shown in Fig.~\ref{fig:results-nonzero-theta}.
Note that at $\theta \ne 0 $,  the G-partiy is no longer the exact symmetry, then we mainly investigate the isospin and parity.
We perform the fit of the data for each meson using the dispersion-relation and obtain the mass $M$ as shown in the left panel in Fig.~\ref{fig:comp-result}.
Here, two data sets of results obtained by the (modified) one-point function scheme and the dispersion-relation scheme are shown simultaneously~\footnote{As for the results of the one-point function scheme, the proceeding~\cite{Matsumoto:2025ivd} gives a brief summary.}.
It can be seen that the $\theta$-dependence of the mass spectra for composite states is clearly obtained, and the methodology dependence is negligible.
Furthermore, the gray dashed and dotted curves denote $M_{\pi}(0) | \cos(\theta/2) |^{2/3}$ and $\sqrt{3} M_{\pi}(0) | \cos(\theta/2) |^{2/3}$, respectively, where they are predicted by the bosonization analyses~\cite{Coleman:1976uz}.

\section{Summary and Outlook}
These days, when first-principles calculations of gauge theory based on Hamiltonian formalism are attracting attention, it is useful to consider how to calculate physical quantities that have been calculated with Lattice Monte Carlo QCD in the Hamiltonian formalism.
In this manuscript, how to evaluate the mass spectra of composite states in the 2-flavor Schwinger model is discussed in the Hamiltonian formalism. 
There are several ways to compute them, and our proposals work well even with the topological $\theta$-term, which induces the sign problem in conventional Monte Carlo methods.
Among them, the direct use of the dispersion relation is compatible with the Hamiltonian formalism and is useful even for rather small volume calculations.

The right panel of Fig.~\ref{fig:comp-result} depicts the result of the Monte Carlo method based on the Lagrangian formula given by H.~Fukaya and T.~Onogi~\cite{Fukaya:2003ph}. In their work, thanks to the improved techniques with the topology-fixed calculation, we can see clear data for the small $\theta$ regime, where the sign problem exists.
However, in the large $\theta$ region, it was hard to obtain the result because of the severe sign problem.
Furthermore, they could not have clear data for the sigma and eta mesons.
On the other hand, in the dispersion-relation scheme, all excited states can be obtained without any theoretical knowledge or improvement from $\theta=0$ calculation.
That is an advantage of this scheme based on the Hamiltonian formula.

In our studies~\cite{Itou:2023img, Itou:2024psm}, the numerical simulations have been done by the DMRG method, but it can be straightforwardly applied for quantum computations if a machine with around 40 (logical) qubits comes in the future.

\acknowledgments
We would like to thank M.~Honda, Y.~Kikuchi, A.~Matsumoto, and Y.~Tanizaki for the fruitful discussions and collaborations. We appreciate A.~Matsumoto for carefully reading the manuscript.
The numerical calculations were carried out on Yukawa-21 at YITP in Kyoto University and the PC clusters at RIKEN iTHEMS.
The work of E.~I. is supported by 
JSPS Grant-in-Aid for Transformative Research Areas (A) JP21H05190, 
JSPS Grant Number JP23H05439 (Kiban (S)), 
JST Grant Number JPMJPF2221 (SQAI),  
JPMJCR24I3 (CREST),  
JSPS Grant Number 25K01001 (Kiban (B)), 
and also supported by Program for Promoting Researches on the Supercomputer ``Fugaku'' (Simulation for basic science: from fundamental laws of particles to creation of nuclei) and (Simulation for basic science: approaching the new quantum era), and Joint Institute for Computational Fundamental Science (JICFuS), Grant Number JPMXP1020230411. 
This work is supported by Center for Gravitational Physics and Quantum Information (CGPQI) at Yukawa Institute for Theoretical Physics.

\bibliographystyle{JHEP}
\bibliography{Nf2_Schwinger.bib, QFT.bib}


\end{document}